\documentclass[onecolumn,showpacs,amsmath,amssymb,11pt]{revtex4}
\input{epsf}

\usepackage{graphicx}
\usepackage{array}
\usepackage{dcolumn,longtable}

\usepackage{rotating,booktabs}
\usepackage{booktabs,threeparttable}

\begin{document}

\title{\bf
The long-range non-additive three-body dispersion interactions for the rare gases, alkali and alkaline-earth atoms}
\author{Li-Yan Tang$^{1}$, Zong-Chao Yan$^{2,3}$, Ting-Yun Shi$^{1}$, James F. Babb$^{4}$ and J. Mitroy$^{5}$}

\affiliation {$^1$ State Key Laboratory of Magnetic Resonance and
Atomic and Molecular Physics, Wuhan Institute of Physics and
Mathematics, Chinese Academy of Sciences, Wuhan 430071, P. R. China}

\affiliation {$^2$ Center for Cold Atom Physics, Chinese Academy of
Sciences, Wuhan 430071, P. R. China}

\affiliation{$^3$ Department of Physics, University of New
Brunswick, Fredericton, New Brunswick, Canada E3B 5A3}

\affiliation {$^{4}$ ITAMP, Harvard-Smithsonian Center for
Astrophysics, Cambridge, Massachusetts 02138, USA}

\affiliation {$^{5}$ School of Engineering, Charles Darwin
University, Darwin NT 0909, Australia}

\date{\today}

\begin{abstract}
The long-range non-additive three-body dispersion interaction
coefficients  $Z_{111}$, $Z_{112}$, $Z_{113}$, and $Z_{122}$ are
computed for many atomic combinations using standard expressions.
The atoms considered include hydrogen, the rare gases, the alkali
atoms (up to Rb) and the alkaline-earth atoms (up to Sr). The term
$Z_{111}$, arising from three mutual dipole interactions is known as
the Axilrod-Teller-Muto coefficient or the DDD
(dipole-dipole-dipole) coefficient. Similarly, the terms $Z_{112}$,
$Z_{113}$, and $Z_{122}$ arise from the mutual combinations of
dipole ($1$), quadrupole ($2$), and octupole ($3$) interactions
between atoms and they are sometimes known, respectively, as DDQ,
DDO, and DQQ coefficients. Results for the four $Z$ coefficients are
given for the homonuclear trimers, for the trimers involving two
like-rare-gas atoms, and for the trimers with all combinations of
the H, He, Li atoms.   An exhaustive compilation of all coefficients
between all possible atomic combinations is presented as
supplementary data.
\end{abstract}

 \maketitle

\section{Introduction}

The DDD (dipole-dipole-dipole) or Axilrod-Teller-Muto interaction
\cite{axilrod43a} is a non-additive dispersion interaction that
arises between three polarizable atoms or molecules. In the
long-range potential energy function it enters at the order of the
inverse ninth power of the characteristic separation of the three
particles.  It is studied in the physics and chemistry of gases,
liquids, adsorption of atoms at surfaces, and materials science, and
more recently for applications to ultra-cold atoms. In addition to
the DDD interaction, there are terms arising from mutual
interactions of higher-than-dipole induced multipoles
\cite{bell70c,doran71a}, such as the dipole-dipole-quadrupole (DDQ),
the dipole-dipole-octupole (DDO), and the
dipole-quadrupole-quadrupole (DQQ) interactions. Values of the
coefficients are useful for creating models of potential energy
surfaces and for comparison with more sophisticated methods that
include exchange energies.

For a variety of reasons, there has been increased interest in the
non-additive three-body dispersion coefficients.  Some of the
primary motivations come from studies of ultra-cold atoms.  For
example photo- and magneto-association techniques are being
developed to form homonuclear and heteronuclear trimers.  The
collisions that govern such events typically occur at low energies
where the importance of dispersion events are enhanced and there
have been a number of studies that have investigated the impact of
three-body non-additive interactions
\cite{soldan03a,cvitas06a,cvitas07a,quemener07a}. Another
application area is in fine-tuning the interaction potentials of
rare gases for the description of their bulk thermophysical
properties such as the virial coefficients
\cite{vanderhoef99a,jakse02a,nasrabad04a,wang06a,jager11a} and for
computing properties of their solid forms \cite{lotrich97a}. There
has also been interest in describing the structures of metal atoms
embedded in rare gases
\cite{baylis77a,niemax80a,sheng09a,tang09a,yin10a}. Finally, the
dispersion interactions give reference data that could be utilized
to validate \textit{ab initio} calculations of the full three-body
potential surfaces at large inter-nuclear separations
\cite{hauser08a,soldan10a,soldan10b}.

The DDD coefficients are known very accurately for systems involving
combinations of H, He and Li through \textit{ab initio} calculations
\cite{stacey68a,stewart73a,koga86a,koga89a,yan96a}, while the most
accurate DDD values for the heavier rare gases are derived from
pseudo-oscillator strength distributions (POSDs) using dipole
polarizabilities and other experimental
data~\cite{kumar85a,kumar85b,standard85a}. Doran and
Zucker~\cite{doran71a} and Doran~\cite{doran72a,doran74a} calculated
higher order coefficients for the rare gases using Hartree-Fock
methods.   Their results were surpassed in accuracy by those of
Thakkar {\em et al}~\cite{thakkar92a}. Reasonably accurate values
for the DDD coefficients were computed for the alkali-metal
\cite{marinescu97a,mitroy03f} and for the alkaline-earth metal atoms
\cite{mitroy03f}.  An alternate approach to the determination of
three-body dispersion coefficients is to use combination rules
\cite{midzuno56a,kihara58a,tang69a,diaz80a}, which utilize estimates
of other properties such as polarizabilities and two-body dispersion
coefficients.

The present work represents the most comprehensive evaluation of
three-body dispersion coefficients so far presented in the
literature. Coefficients are computed using oscillator strength sum
rules up to a combined inverse power of 15 in the interspecies
separation distance are given.  Extremely precise dispersion
coefficients are given for all combinations of the H, He and Li
ground states since definitive results for these atoms can often
serve as benchmarks to test methodologies used for more complicated
systems. In addition, dispersion coefficients are given for all the
three-body homonuclear trimers involving the rare gases, the alkali
atoms up to Rb, and the alkaline earth atoms up to Sr. Space
requirements preclude the presentation of data for all possible
trimers involving heteronuclear systems.  Tabulated coefficients are
only given for systems with two like rare gas atoms. The three-body
dispersion coefficients for almost all other possible three-body
systems are detailed in the supplementary data submitted with this
paper.

\section{Definitions of the dispersion coefficients}

\subsection{Expressions using oscillator strength sum rules}

The leading non-additive dipole-dipole-dipole dispersion interaction
between three atoms (a, b, c) can be written as
\begin{equation}
V^{\rm abc}_{111}(\mathbf{R}_{\rm a},\mathbf{R}_{\rm
b},\mathbf{R}_{\rm c}) = Z^{\rm abc}_{111} W^{\rm
abc}_{111}(\mathbf{R}_{\rm a},\mathbf{R}_{\rm b},\mathbf{R}_{\rm c})
\,,
\end{equation}
where  $Z^{\rm abc}_{111}$  is  the three-body dispersion
coefficient  and the function $W^{\rm abc}_{111}(\mathbf{R}_{\rm
a},\mathbf{R}_{\rm b},\mathbf{R}_{\rm c})$ is detailed in
\cite{axilrod43a,cvitas06a}, e.g.
\begin{equation}
W^{\rm abc}_{111}(\mathbf{R}_{\rm a},\mathbf{R}_{\rm b},\mathbf{R}_{\rm c})
 = \frac{ 1 + 3 \cos(\theta_{\rm ab})\cos(\theta_{\rm bc})\cos(\theta_{\rm ca}) }
{R^3_{\rm ab}R^3_{\rm bc}R^3_{\rm ca}}  \,, \label{WDDD}
\end{equation}
where $\theta_{\rm ab}$ is the interior angle at $c$ due to atoms $a$
and $b$, $R_{\rm ab}$ is the distance between
atoms $a$ and $b$, and
\begin{eqnarray}
Z^{\rm abc}_{111} &=& \frac{3}{2} \sum_{ijk}
\frac{ f^{(1)}_{{\rm a},0i} f^{(1)}_{{\rm b},0j} f^{(1)}_{{\rm c},0k} }
{ \epsilon_{{\rm a},0i} \epsilon_{{\rm b},0j} \epsilon_{{\rm c},0k} } \nonumber \\
  & \times & \frac{ (\epsilon_{{\rm a},0i}+\epsilon_{{\rm b},0j}+\epsilon_{{\rm c},0k})}
{ (\epsilon_{{\rm a},0i}+\epsilon_{{\rm b},0j}) (\epsilon_{{\rm
a},0i}+\epsilon_{{\rm c},0k})(\epsilon_{{\rm b},0j}+\epsilon_{{\rm
c},0k})} \,. \label{ZDDD}
\end{eqnarray}
The sums over $i, j,$ and $k$ involve the dipole oscillator
strengths $f^{(1)}_{0i}$ and the energy difference $\epsilon_{0i}$
from the ground state of each atom. There are alternate definitions
of $Z_{111}$ where $Z_{111}$ is a factor of 3 smaller with the
$W_{111}$ geometric factors having an extra multiplying factor of 3
\cite{bell70c,doran74a,thakkar92a}.

More generally, the three-body potential for arbitrary multipole
combinations can be written
\begin{equation}
V^{\rm abc}_{\ell_a \ell_b \ell_c}(\mathbf{R}_{\rm
a},\mathbf{R}_{\rm b},\mathbf{R}_{\rm c}) = Z^{\rm abc}_{\ell_a
\ell_b \ell_c} W^{\rm abc}_{\ell_a \ell_b \ell_c} (\mathbf{R}_{\rm
a},\mathbf{R}_{\rm b},\mathbf{R}_{\rm c}) \,,
\end{equation}
where the general dispersion coefficient is written
\begin{eqnarray}
Z^{abc}_{\ell_a \ell_b \ell_c} &=& \frac{3}{2} \sum_{ijk}
\frac{ f^{(\ell_a)}_{{\rm a},0i} f^{(\ell_b)}_{{\rm b},0j} f^{(\ell_c)}_{{\rm c},0k} }
{ \epsilon_{{\rm a},0i} \epsilon_{{\rm b},0j} \epsilon_{{\rm c},0k} } \nonumber \\
  & \times & \frac{ (\epsilon_{{\rm a},0i}+\epsilon_{{\rm b},0j}+\epsilon_{{\rm c},0k})}
{ (\epsilon_{{\rm a},0i}+\epsilon_{{\rm b},0j}) (\epsilon_{{\rm
a},0i}+\epsilon_{{\rm c},0k})(\epsilon_{{\rm b},0j}+\epsilon_{{\rm
c},0k})} \,.  \label{C9}
\end{eqnarray}
The $f^{(\ell)}_{0i}$ refer to the multipole oscillator strength
from the ground state of each atom according to the definitions in
\cite{yan96a,mitroy03f}, e.g.
\begin{equation}
f^{(\ell)}_{0i}=\frac{2}{(2\ell+1)}\epsilon_{0i}|\langle
\Psi_0||\sum_j r_j^{\ell}{\mathbf
C}_{\ell}(\hat{\mathbf{r}}_j)||\Psi_i \rangle \big |^2 \,.
\end{equation}

The next term in the three-body dispersion  series is the
dipole-dipole-quadrupole (DDQ) contribution. In the case of a
heteronuclear trimer there are three separate terms since the
quadrupole excitation can be associated with any of the atoms, $a$,
$b$, or $c$. One term is
\begin{equation}
V^{\rm abc}_{112}(\mathbf{R}_{\rm a},\mathbf{R}_{\rm
b},\mathbf{R}_{\rm c}) = Z^{\rm abc}_{112} W^{\rm
abc}_{112}(\mathbf{R}_{\rm a},\mathbf{R}_{\rm b},\mathbf{R}_{\rm c})
\,. \label{V3body}
\end{equation}
The $Z^{\rm abc}_{112}$ coefficient is
\begin{eqnarray}
Z^{\rm abc}_{112} &=& \frac{3}{2} \sum_{ijk}
\frac{ f^{(1)}_{{\rm a},0i} f^{(1)}_{{\rm b},0j} f^{(2)}_{{\rm c},0k} }
{ \epsilon_{{\rm a},0i} \epsilon_{{\rm b},0j} \epsilon_{{\rm c},0k} } \nonumber \\
  & \times & \frac{ (\epsilon_{{\rm a},0i}+\epsilon_{{\rm b},0j}+\epsilon_{{\rm c},0k})}
{ (\epsilon_{{\rm a},0i}+\epsilon_{{\rm b},0j}) (\epsilon_{{\rm
a},0i}+\epsilon_{{\rm c},0k})(\epsilon_{{\rm b},0j}+\epsilon_{{\rm
c},0k})} \,. \label{Z112}
\end{eqnarray}
The spatial factor (dropping the arguments for compactness) for
$W^{\rm abc}_{112}$ is
\begin{eqnarray}
W^{\rm abc}_{112}
 &=& \frac{ 1 } {16 R^3_{\rm ab}R^4_{\rm bc}R^4_{\rm ca} } \times
 \bigg[ 9 \cos(\theta_{\rm ab}) - 25 \cos(3\theta_{\rm ab})  \nonumber \\
  &+& 6 \cos(\theta_{\rm bc}\!-\!\theta_{\rm ca})
  [3+5\cos(2\theta_{\rm ab})] \bigg] \,.
\label{WDDQ}
\end{eqnarray}
There is no need to give the $Z$ coefficients for the other combinations.
The spatial factors for $W^{\rm abc}_{122}$, $W^{\rm abc}_{113}$ and
$W^{\rm abc}_{222}$ can be written \cite{bell70c,cvitas06a} as
\begin{eqnarray}
W^{\rm abc}_{122}\! &=& \! \frac{ 5 } {64 R^4_{\rm ab}R^5_{\rm
bc}R^4_{\rm ca} }
\! \times \! \bigg[3 \cos(\theta_{\rm bc}) + 15 \cos(3\theta_{\rm bc})  \nonumber \\
 \! &+& \! 20 \cos(\theta_{\rm ca}\!-\!\theta_{\rm ab})
  [1-3\cos(2\theta_{\rm bc})] \nonumber \\
 \! & + & \! 70\cos[2(\theta_{\rm ca}-\theta_{\rm ab})] \cos(\theta_{\rm bc})
  \bigg] \,,   \\
W^{\rm abc}_{113} \!
 &=& \! \frac{ 5 } {96 R^3_{\rm ab}R^5_{\rm bc}R^5_{\rm ca} } \! \times \!
 \bigg[ 9 + 8\cos(2\theta_{\rm ab}) - 49 \cos(4\theta_{\rm ab})  \nonumber \\
  &+& 6 \cos(\theta_{\rm bc}\!-\theta_{\rm ca}\!)\big[9\cos\theta_{\rm ab}+7\cos(3\theta_{\rm ab})\big] \bigg]
\label{WDDO} \,,
\end{eqnarray}
\begin{eqnarray}
W^{\rm abc}_{222} &=& \frac{ 5 } {128 R^5_{\rm ab}R^5_{\rm
bc}R^5_{\rm ca} } \! \times \! \bigg[
 490 \cos(2\theta_{\rm ab})\cos(2\theta_{\rm bc})\cos(2\theta_{\rm ca}) \nonumber \\
 & - & 27 + 220 \cos(\theta_{\rm ab})\cos(\theta_{\rm bc})\cos(\theta_{\rm ca}) \nonumber \\
  &+& 175 \bigg\{ \cos[2(\theta_{\rm bc}\!-\!\theta_{\rm ca})] +
   \cos[2(\theta_{\rm ca}\!-\!\theta_{\rm ab})] \nonumber \\
  &+& \cos[2(\theta_{\rm ab}\!-\!\theta_{\rm bc})]  \bigg\} \bigg]
  \,.
\label{WQQQ}
\end{eqnarray}
Expressions for rearrangements of the different multipole
excitations amongst the different atoms can be written down
transparently. The expressions for the homonuclear trimers, i.e.
$Z^{\rm aaa}_{\ell_1 \ell_2 \ell_3}$ and $W^{\rm aaa}_{\ell_1 \ell_2
\ell_3}$ do not require any modifications to the formulae listed
above.

\subsection{Combination rules}

The present approach to the calculation of the dispersion parameters
is based on explicit evaluation of oscillator strength sum rules. An
alternative approach is to use integrals over the polarizabilities
at imaginary frequencies \cite{yan96a,cvitas06a}.  Approximate
expressions that use existing calculations of polarizabilities and
two-body dispersion parameters in combination rules have been
derived \cite{midzuno56a,kihara58a,bell70c}.  One of these
combination rules is known as the Midzuno-Kihara approximation
\cite{midzuno56a,kihara58a}. For a trimer consisting of three
identical atoms, one has
\begin{equation}
C_9 =  \frac{3}{4} \alpha_d C_6  \label{C9MK}
\end{equation}
Other more complicated expressions exist for
heteronuclear trimers \cite{midzuno56a,kihara58a} and   other
combination rules have been developed \cite{diaz80a}.  Estimates
using the Midzuno-Kihara approximation for $Z_{111}$ are given
in some tables and abbreviated as MK.

\begin{table*}
\caption{\label{XXX} The dispersion constants (in a.u.) for
homonuclear trimers consisting of either hydrogen, helium or
lithium.  All other dispersion coefficients are given by the
symmetry conditions as described in the text. The numbers in
parentheses are the computational uncertainties arising from
incomplete convergence of the basis set. Values for hydrogen are
correct to all quoted digits. The numbers in the square brackets
denote powers of ten. }
\begin{ruledtabular}
\begin{tabular}{lccccc}
\multicolumn{1}{l}{System}  & \multicolumn{1}{c}{$Z_{111}$} &
\multicolumn{1}{c}{$Z_{112}$}
& \multicolumn{1}{c}{$Z_{122}$} &  \multicolumn{1}{c}{$Z_{113}$} & \multicolumn{1}{c}{$Z_{222}$} \\
\hline
H      &  21.642464510636   & 78.707166385334    & 289.37119056926   & 712.03297892387    & 1077.4737895329  \\
He   &  1.47955860643(1)    & 2.77280376179(1)   & 5.21831397022(6)  & 12.37878701399(5)  & 9.8654795914(2)\\
Li   &  1.70615(2)[5]       & 1.78338(2)[6]      & 1.99321(1)[7]     & 5.12903(4)[7]      & 2.463549(6)[8]\\
\end{tabular}
\end{ruledtabular}
\end{table*}

\section{Underlying description of atomic structures}

The evaluation of the three-body dispersion parameters involves the
construction of tables of oscillator strengths to describe the
excitation spectrum of all atoms considered.  These oscillator
strength distributions were constructed by a variety of means
depending on the specifics of each atom.  A short description is now
given for each atom since the accuracy of the dispersion
coefficients depends on the details of the underlying structure
model.

\subsection{Hydrogen}

The hydrogen atom and its excitation spectrum was computed by
diagonalising the hydrogen hamiltonian in a basis of Laguerre type
orbitals (LTOs) \cite{yan99c,mitroy05a}. Such a basis has the
property that it can be expanded to completeness without any linear
dependence issues. Very accurate dispersion parameters can be
obtained with relatively small basis sets.

\subsection{Hylleraas descriptions for He and Li}

The helium atom wave function is expanded as a linear combination of
Hylleraas functions.  The resulting oscillator strength sum rules
are capable of giving the dipole polarizability to an accuracy of 12
significant figures.  The lithium atom wave function can also be
represented as a linear combination of Hylleraas functions.  The
achievable accuracy for the Li ground state dipole polarizability is
5 significant figures. Details of the Hylleraas method and its
application to He and Li can be found in Refs \cite{yan96a,tang09a}.

\subsection{Pseudo-oscillator strength distributions for heavier rare gases}

The pseudo-oscillator strength distributions  (POSDs) for the
heavier rare gases, Ne, Ar, Kr and Xe, come from two sources.  The
dipole distributions are taken from Kumar and Meath
\cite{kumar85a,kumar85b}, which were constructed by reference to
polarizability data and other experimental data.  The quadrupole and
octupole pseudo-oscillator strengths are based on Hartree-Fock (HF)
(or relativistic HF) single particle energies and radial expectation
values. These were given tunings to reproduce calculations of higher
multipole polarizabilities and homonuclear dispersion coefficients
\cite{mitroy07d}.

\subsection{Semi-empirical description of structure for alkali and alkaline-earth atoms}

The transition arrays for the alkali atoms are those which were used
in calculations of the dispersion interactions between these atoms
and the ground states of hydrogen and helium \cite{zhang07c}. The
transition arrays for the alkaline-earth atoms are those which were
used in calculations of the dispersion interactions involving the
alkali- and alkaline-earth atoms \cite{mitroy03f}.

The wave function is partitioned into core and valence parts, where
the core is described by a HF wave function.  The ground and excited
state pseudo-spectrum are computed by diagonalizing the fixed-core
Hamiltonian of one and two electron functions written as a linear
combination of LTOs.  The HF core Hamiltonian is supplemented with a
semi-empirical core polarization potential tuned to reproduce the
energies of the low-lying spectrum. For all practical purposes the
results are converged with respect to increasing the dimension of
the orbital basis.

Excitations from the core are included in the calculations of the
dispersion coefficients by the construction of pseudo-oscillator
strength distributions based on HF energies and expectation values
\cite{mitroy03f}. This approach to the determination of atomic
structure is referred to as the configuration interaction plus core
polarization (CICP) method in the remainder of this article.

\begin{table*}
\caption{\label{XXY}The three atom dispersion constants for
combinations of H, He and Li with two like atoms.  The other
dispersion constants are given by the symmetry relations $Z^{\rm
aab}_{211}$=$Z^{\rm aab}_{121}$, $Z^{\rm aab}_{212}$=$Z^{\rm
aab}_{122}$, and $Z^{\rm aab}_{311}$=$Z^{\rm aab}_{131}$ (in a.u.).
The numbers in parentheses are the computational uncertainties
arising from the finite basis set size. }
\begin{ruledtabular}
\begin{tabular}{lcccc}
\multicolumn{1}{l}{System}  & \multicolumn{1}{c}{$Z^{\rm aab}_{111}$} &
\multicolumn{1}{c}{$Z^{\rm aab}_{112}$} & \multicolumn{1}{c}{$Z^{\rm aab}_{121}$}
& \multicolumn{1}{c}{$Z^{\rm aab}_{122}$}\\
\hline
H-H-He      &  8.10224087465(2) & 14.6777067844(1) & 30.3021391145(1)  & 55.1425238705(2)\\
H-H-Li      &  275.987(3)       & 4195.023(6)      & 944.97(1)         & 14670.29(2)\\
He-He-H     &  3.26806489632(2) & 12.7469884056(1) & 5.99712674013(3)  & 23.5605644280(1)\\
He-He-Li    &  29.8259(6)       & 504.658(1)       & 53.228(1)         & 907.467(2)  \\
Li-Li-H     &  6133.7(1)        & 20621.1(2)       & 72048.9(6)        & 243884(2)\\
Li-Li-He    &  1917.48(2)       & 3394.56(4)       & 22845.5(2)        & 40500.9(4)\\
\hline
\multicolumn{1}{l}{System}  & \multicolumn{1}{c}{$Z^{\rm aab}_{221}$} &  \multicolumn{1}{c}{$Z^{\rm aab}_{113}$} &  \multicolumn{1}{c}{$Z^{\rm aab}_{131}$} & \multicolumn{1}{c}{$Z^{\rm aab}_{222}$} \\
\hline
H-H-He      & 115.1382084748(3)& 64.4548538033(2)    & 277.5076627177(7) & 210.661799531(1)\\
H-H-Li      & 3245.34(5)       & 135490.15(9)        & 8349.17259(1)     & 51577.89(8)\\
He-He-H     & 11.0328224804(1) & 118.9339509877(6)   & 26.4993528199(1)  & 43.6808551603(3)\\
He-He-Li    & 95.057(2)        & 17139.665(9)        & 232.194(6)        & 1633.531(4)\\
Li-Li-H     & 966692(2)        & 180949(2)           & 214091(2)         & 3312094(6)\\
Li-Li-He    & 314438.8(7)      & 14753.5(2)          & 682925(7)         & 558901(2)\\
\end{tabular}
\end{ruledtabular}
\end{table*}

\section{Numerical Results}

\subsection{H, He and Li}

A complete set of dispersion coefficients for all combinations of
the light systems involving H, He and Li up to a total inverse power
of $R^{15}$ are given in Tables \ref{XXX}, \ref{XXY} and \ref{XYZ}.
The dispersion coefficients for these systems are given special
prominence since they can be used to validate high precision quantum
chemistry calculations of potential energy surfaces for these
trimers. Table \ref{HHeLi} compares values of $Z_{111}$ with CICP
results to give an accuracy benchmark for the CICP calculations.

Certain symmetry conditions apply when the trimer consists of three
homonuclear atoms.  One can write $Z^{\rm aaa}_{121}$=$Z^{\rm
aaa}_{211}$=$Z^{\rm aaa}_{112}$, $Z^{\rm aaa}_{212}$=$Z^{\rm
aaa}_{221}$=$Z^{\rm aaa}_{122}$, and $Z^{\rm aaa}_{131}$=$Z^{\rm
aaa}_{311}$=$Z^{\rm aaa}_{113}$. The results for the H, He and Li
homonuclear trimers are given in Table \ref{XXX}.  The dispersion
coefficients for hydrogen are given to 14 digits.  This does not
represent the achievable precision.  It is possible to compute the
$Z_{\ell_1 \ell_2 \ell_3}$ to 30 significant digits by using
quadruple precision arithmetic and a basis of 30 Laguerre functions.
The $Z_{\ell_1\ell_2\ell_3}$ coefficients are in perfect agreement
with those of Koga \cite{koga89a}.  The Koga values use different
conventions for relating the $Z_{\ell_1\ell_2\ell_3}$ to the
geometric functions and are not quoted in Table \ref{homo}.

For helium, a precision of at least 10 significant digits was
achieved, while for lithium six digits were tabulated with some
uncertainty usually occurring in the sixth digit.  There have been
no previous high precision calculations of these coefficients except
for $Z_{111}$ which were previously computed by Yan {\em et al}
using the Hylleraas basis \cite{yan96a}. The present coefficients
are in harmony with these earlier calculations.

The uncertainties in the dispersion parameters were estimated by an
examination of the convergence for a series of successively larger
calculations \cite{tang09a,yan96a}. Relativistic and finite mass
corrections for these atoms can be expected to start to occur by the
fourth to sixth digits. Therefore the precision achieved by the
present non-relativistic calculations is sufficient for any
practical calculation since finite-mass and relativistic corrections
would need to be incorporated to further improve the overall level
of accuracy.

\begin{table}
\caption{\label{XYZ}The dispersion coefficients for the heteronuclear
H-He-Li trimer (in a.u.). The numbers in parentheses are the computational
uncertainties arising from incomplete convergence of the basis set.}
\begin{ruledtabular}
\begin{tabular}{lcccccccccc}
\multicolumn{1}{c}{$Z_{\ell_1 \ell_2 \ell_3}$} &
\multicolumn{1}{c}{Values}  &
\multicolumn{1}{c}{$Z_{\ell_1\ell_2\ell_3}$} &
\multicolumn{1}{c}{Values}
 & \multicolumn{1}{c}{$Z_{\ell_1\ell_2\ell_3}$} &  \multicolumn{1}{c}{Values}   \\
 \hline
$Z_{111}$  & 89.833(1)    &  $Z_{112}$ & 1430.195(3)    &  $Z_{121}$  & 159.785(3)  \\
$Z_{211}$  & 309.922(6)   &  $Z_{122}$ & 2557.314(4)    &  $Z_{212}$  & 5068.475(9)  \\
$Z_{221}$  & 551.89(1)    &  $Z_{113}$ & 47174.62(3)    &  $Z_{131}$  & 695.90(1)  \\
$Z_{311}$  & 2747.29(5)   &  $Z_{222}$  & 9081.12(2) \\
\end{tabular}
\end{ruledtabular}
\end{table}

The $Z_{111}$ coefficients for all H, He, Li combinations are given
in Table \ref{HHeLi}.  They are compared with the earlier Hylleraas
calculations of Yan {\em et al} \cite{yan96a} and coefficients from
from CICP type calculations \cite{mitroy03f}. The two sets of
Hylleraas for trimers containing just H and He are generally in
agreement when the quoted uncertainties are taken into
consideration.  However, the present dispersion coefficients for
helium are about one order of magnitude more precise than the
calculations by Yan {\em et al} \cite{yan96a}.

In a  few instances involving the Li atom the present $Z_{111}$
values lie outside the error bars of the earlier Hylleraas
calculation. The basis sets used in the Yan {\em et al} calculation
were much smaller than those used herein.  Agreement between the
CICP and Hylleraas calculations is good, with no differences
exceeding $0.2\%$. The older $Z_{111}$ values of
\cite{bell70c,bell70d} and \cite{stacey68a} are not listed since
they are not expected to have the precision of the present
calculations.

\begin{table}
\caption{\label{HHeLi} Comparison of the $Z_{111}$ (in a.u.) parameter for all
combinations of the trimers formed by the H, He, and Li with some
earlier calculations.  The numbers in parentheses are the computational
uncertainties arising from incomplete convergence of the basis set.}
\begin{ruledtabular}
\begin{tabular}{lccc}
\multicolumn{1}{c}{ } & \multicolumn{2}{c}{Hylleraas}  & \multicolumn{1}{c}{CICP} \\
\multicolumn{1}{c}{System} & \multicolumn{1}{c}{Present}  &
\multicolumn{1}{c}{Ref. \cite{yan96a}} & \multicolumn{1}{c}{Ref. \cite{mitroy03f}} \\
\hline
H-H-H      & 21.64246451063597        & 21.642464510635  & 21.6425  \\
H-H-He     & 8.10224087465(2)         & 8.1022408743(2)  & 8.103  \\
H-H-Li     & 275.987(3)               & 275.979(7)       & 276.0  \\
He-He-He   & 1.47955860643(1)         & 1.4795586063(1)  & 1.4798  \\
He-He-H    & 3.26806489632(2)         & 3.2680648961(1)          & 3.269  \\
He-He-Li   & 29.8259(6)               & 29.824(5)                & 29.83  \\
Li-Li-Li   & 170615(2)                & 170595(6)                & 170873 \\
Li-Li-H    & 6133.7(1)                & 6133.5(5)                 & 6139  \\
Li-Li-He   & 1917.48(2)               & 1917.27(5)               & 1919  \\
Li-He-H    & 89.833(1)                & 89.830(5)                & 89.85  \\
\end{tabular}
\end{ruledtabular}
\end{table}

The relative importance of the Z$_{111}$ coefficient with respect to
the two-body dispersion coefficients is easy to estimate and the
Li-Li-Li trimer is used as an example.  The two body $C_8$ and
$C_{10}$ dispersion coefficients are $C_8=8.3429(1)\times10^4$ a.u.
and $C_{10}=7.3725(2)\times 10^6$ a.u. \cite{tang09a}.  When three
Li atoms form an equilateral triangle the spatial factor of
Eq.~(\ref{WDDD}) is $\frac{11}{8}R^{-9}$.  At $R = 30$ $a_0$ the
ratios of $Z_{111}W_{111}$ to $C_8 R^{-8}$ and $C_{10}R^{-10}$ are
0.28 and 0.95 respectively.  The nonadditive three-body interaction
should be included in high-precision analysis of photo-association
spectra for homonuclear alkali-metal trimers. These ratios are
0.0096 and 0.1661 respectively for a helium trimer in an equilateral
triangle configuration at $R = 15$ $a_0$.

\begin{table*}[th]
\caption[]{  \label{homo} The three-body dispersion coefficients,
$Z_{111}$, $Z_{112}$, $Z_{122}$ and $Z_{113}$ (in atomic units) for
homonuclear trimers.  The $f$-value distributions for H, He and Li
use Laguerre type orbitals (H) or Hylleraas basis functions (He and
Li) to describe the ground and excited state spectra.  Dispersion
coefficients for the H, He and Li trimers are given to additional
significant digits in Table \ref{XXX}. The heavier gas $f$-value
distributions use pseudo-oscillator strength distributions
\cite{kumar85a,kumar85b,mitroy07b}, while those for the other atoms
come from CICP calculations.   Results in the Midzuno-Kihara (MK)
approximation \cite{midzuno56a,kihara58a} use the present oscillator
strength distributions to compute the underlying $\alpha_d$ and
$C_6$ needed as input. The numbers in the square brackets denote
powers of ten. } \vspace{0.1cm}
\begin{ruledtabular}
\begin{tabular}{lccccccc}
Systems &  \multicolumn{3}{c}{$Z_{111}$} &
          \multicolumn{2}{c}{$Z_{112}$}  &
          \multicolumn{1}{c}{$Z_{122}$}  &
          \multicolumn{1}{c}{$Z_{113}$}  \\
\cline{2-4} \cline{5-6}
     & Present & Other  &  MK       &  Present &  Other   \\ \hline
He   &  1.4796     & 1.47956 \footnotemark[1]    & 1.516       &  2.77280    & 2.77280 \footnotemark[2]   & 5.2218     & 12.378  \\
Ne   & 11.95       & 12.37  \footnotemark[3]     & 12.78       &  33.73      & 34.20 \footnotemark[4]     & 95.21      & 189.7 \\
Ar   & 518.3       & 521.7 \footnotemark[3]      & 534.4       &  2.543[3]   & 2.554[3] \footnotemark[1]  & 1.251[4]   & 2.535[4]  \\
Kr   & 1.572[3]    & 1.697[3] \footnotemark[3]   & 1.631[3]    &  9.556[3]   & 10.25[3] \footnotemark[3]  & 5.827[4]   & 1.174[5] \\
Xe   & 5.573[3]    &  6.246[3] \footnotemark[3]  & 5.822[3]    &  4.582[4]   & 5.163[4] \footnotemark[3]  & 3.778[5]   & 7.298[5] \\
H    & 21.624      &  21.642 \footnotemark[2]    & 21.93       &  78.707     & 78.72 \footnotemark[2]     & 289.37     & 712.03    \\
Li   & 1.7062[5]   & 1.701[5] \footnotemark[4]   & 1.717[5]    &  1.786[6]   & 1.750[6] \footnotemark[4]  & 1.997[7]   & 5.138[7]  \\
Na   & 1.892[5]    & 1.906[5] \footnotemark[4]   & 1.908[5]    &  2.544[6]   & 2.353[6] \footnotemark[7]  & 3.553[7]   & 7.844[7] \\
K    & 8.318[5]    & 8.493[5] \footnotemark[4]   & 8.488[5]    &  1.645[7]   & 1.493[7] \footnotemark[7]  & 3.354[8]   & 6.212[7]   \\
Rb   & 1.063[6]    & 1.097[5] \footnotemark[4]   & 1.097[6]    &  2.455[8]   & 2.230[8] \footnotemark[7]  & 5.678[8]   & 9.704[8]   \\
Be   & 5.973[3]    & 6.135[3] \footnotemark[5]   & 6.023[3]    &  5.335[5]   & 5.535[4]  \footnotemark[5] & 4.865[5]   & 7.420[5]        \\
Mg   & 3.338[4]    & 3.36[4] \footnotemark[5]    & 3.369[3]    &  4.182[5]   & 4.26[5] \footnotemark[5]   & 5.307[6]   & 7.667[6]          \\
Ca   & 2.557[5]    & 3.25[5] \footnotemark[5]    & 2.626[5]    &  5.131[6]   & 5.34[6] \footnotemark[5]   & 1.034[8]   & 1.125[8]   \\
Sr   & 4.753[5]    & 4.903[5] \footnotemark[6]   & 4.903[5]    &  1.111[7]   &                            & 2.601[8]   & 2.983[8]   \\
\end{tabular}
\end{ruledtabular}
\footnotetext[1]{Independent Hylleraas calculation \cite{yan96a}}.
\footnotetext[2]{Pseudo oscillator strength distribution based on
accurate oscillator strength sum rules \cite{bell70d}}
\footnotetext[3]{Many body perturbation theory (MBPT)
\cite{thakkar92a}} \footnotetext[4]{Model potential
\cite{marinescu97a}} \footnotetext[5]{Pseudo oscillator strength
distributions (POSD). Value given is mean of the lower and upper
limits. \cite{standard85a}} \footnotetext[6]{MBPT \cite{doran74a}}
\footnotetext[7]{Asymptotic wave function \cite{huang11a}}
\end{table*}

\subsection{Homonuclear trimers}

Dispersion coefficients for the homonuclear trimers are given in
Table \ref{homo}. The atoms presented are hydrogen, the rare gases,
the alkali atoms, and the alkaline earth atoms.  Values of $Z_{111}$
for most of these atoms have been published previously
\cite{stacey68a,langhoff70a,stewart73a,doran74a,margoliash78a,standard85a,thakkar92a,yan96a,marinescu97a,patil97a,mitroy03f,huang11a}.

The set of values for the rare gases heavier than He should be
regarded as the current benchmark. The $Z_{111}$ coefficients use
the pseudo oscillator strength distributions of Kumar and Meath
\cite{kumar85a,kumar85b} and should be accurate to about $\pm 1\%$.
The best \textit{ab initio} calculation is the MBPT calculation
\cite{thakkar92a} which should be regarded as superseding earlier
calculations \cite{bell66b,tang71a,doran71a,doran72a,doran74a}. The
MBPT calculation overestimates the dipole polarizabilities of Kr and
Xe and this leads to the MBPT $Z_{111}$ values being too large for
these atoms.

The present $Z_{112}$ values are slightly smaller than the MBPT values.  The present
values should be adopted as the recommended values for reasons outlined above since
the most important multipole excitation is still the dipole and the pseudo oscillator
strength distribution of Kumar and Meath is still the preferred value.   The
MBPT calculation \cite{thakkar92a} also gave values for $Z_{122}$ and $Z_{113}$.
These are not tabulated for reasons of brevity since they repeat the same pattern seen
for the $Z_{111}$ and $Z_{112}$ coefficients.  The MBPT are about 5--10$\%$ larger
than the present values.

A detailed comparison of the CICP alkali atom $Z_{\rm 111}$
coefficients with those of a model potential calculation
\cite{marinescu97a} listed in Table \ref{homo} has been made
previously \cite{mitroy03f}. The CICP $Z_{\rm 111}$ coefficients
were to be preferred since the model potential calculation
\cite{marinescu97a} did not properly incorporate the contributions
from the core.  Two-body dispersion coefficients from the CICP
calculations reproduce high quality relativistic many-body
perturbation theory estimates to a high degree of precision
\cite{mitroy03f,derevianko99a,safronova99a,derevianko01a,porsev03a}.

For the alkali atoms there have also been estimates of the
three-body dispersion coefficients based on a simple model potential
tuned to give the correct binding energies for the lowest state in
each angular momentum \cite{patil97a,huang11a}.   The $Z_{\rm 112}$
results of Huang and Sun \cite{huang11a} are listed in Table
\ref{homo}.  The Huang and Sun $Z_{112}$ tend to be 5-10$\%$ smaller
than the present CICP values.  Taking the Na-Na-Na trimer as a
specific case, the Huang and Sun $Z_{112}$ of $2.353\times 10^6$
a.u. value is 7$\%$ smaller than the present value.  A similar
situation also occurs for the two-body $C_8$ coefficient for the
Na-Na dimer.  The CICP calculations gives $1.159 \times 10^5$ a.u.
\cite{mitroy03f,mitroy05b} while the Huang and Sun calculation gives
$1.083 \times 10^5$ a.u.  There are similar discrepancies with the
Patil and Tang \cite{patil97a} tabulation.  Patil and Tang give
$2.35\times 10^6$ a.u. for the Na-Ne-Ne $Z_{112}$, which is $7\%$
smaller than the present value. The present CICP calculations expose
the limitations of the structure models underlying the Huang and Sun
\cite{huang11a} and Patil and Tang \cite{patil97a} tabulations. The
Patil and Tang three-body coefficients, $Z_{111}$ and $Z_{112}$ were
used as input in an attempt to make a global triatomic potential
surface for three lithium atoms \cite{cvitas06a}.

The present CICP $Z_{\ell_1 \ell_2 \ell_3}$ dispersion coefficients
should be regarded as the reference values.  For example, the
$Z_{111}$ values from the Standard and Certain compilation
\cite{standard85a} listed in Tables \ref{homo} are taken as the mean
values of the lower and upper bounds. The oscillator strength
distributions for some of the atoms in the Standard and Certain
compilation rely on much older and smaller calculations.

\subsection{Impurity atoms embedded in the rare gases}

The three-body dispersion coefficients for the X-He-He systems given
in Table \ref{XHeHe}, where X denotes H, the alkali atoms up to Rb
and the alkaline earth atoms up to Sr. The quadrupole excitation
occurs on one of the He atoms for the $Z^{\rm XHeHe}_{112} = Z^{\rm
XHeHe}_{121}$ coefficients. The quadrupole excitation is associated
with the impurity X atom for the $Z^{\rm XHeHe}_{211}$ coefficient.

\begin{table*}[th]
\caption[]{  \label{XHeHe} The three-body dispersion coefficients
(in atomic units), for the X-He-He systems containing two helium
atoms.  The sources for the $f$-values distributions are the same as
those in Table \ref{homo}.  The dispersion coefficients involving H
and Li are given with additional significant digits in Table
\ref{XXY}. Results in the Midzuno-Kihara (MK) approximation
\cite{midzuno56a,kihara58a} use the present oscillator strength
distributions to compute the underlying $\alpha_d$ and $C_6$ needed
as input. The numbers in the square brackets denote powers of ten.
} \vspace{0.1cm}
\begin{ruledtabular}
\begin{tabular}{lcccccccc}
Systems &  \multicolumn{2}{c}{$Z_{111}$} &
          \multicolumn{1}{c}{$Z_{112}$}  &  \multicolumn{1}{c}{$Z_{211}$}  &
          \multicolumn{1}{c}{$Z_{122}$}  &  \multicolumn{1}{c}{$Z_{221}$}  &
          \multicolumn{1}{c}{$Z_{113}$}  &  \multicolumn{1}{c}{$Z_{311}$}  \\
\cline{2-3}
    &     Present & MK &      \\ \hline
Ne   & 2.955  &  3.067   &  5.561  &  8.335  & 10.51   & 15.69  & 24.86   & 46.96  \\
Ar   & 10.25  & 10.52    &  19.04  & 50.99   & 35.51   & 91.11  & 84.65   & 505.1   \\
Kr   & 14.62  & 15.05    &  27.09  & 90.66   & 50.37   & 168.6  & 120.3   & 1099   \\
Xe   & 21.74  & 22.46    &  40.15  & 183.2   & 74.35   & 339.4  & 177.9   & 2884  \\
H    & 3.269  & 3.324    &  5.999  & 12.75   & 11.04   & 23.57  & 26.52   & 118.9   \\
Li   & 29.83  & 30.13    &  53.25  & 505.1   & 95.11   & 908.5  & 232.6   & 1.716[4]  \\
Na   & 33.52  & 34.30    &  59.99  & 624.8   & 107.5   & 1123   & 262.4   & 2.279[4]    \\
K    & 50.73  & 52.82    &  90.97  & 1276    & 163.4   & 2287   & 398.1   & 6.014[4]  \\
Rb   & 56.80  & 59.75    &  106.0  & 1538    & 183.6   & 2753   & 447.0   & 7.676[4]  \\
Be   & 16.54  & 16.77    &  29.90  & 176.3   & 54.13   & 321.2  & 131.3   & 2749   \\
Mg   & 27.05  & 27.56    &  48.78  & 392.2   & 88.07   & 710.1  & 213.9   & 8249    \\
Ca   & 46.34  & 47.87    &  83.42  & 1008    & 150.4   & 1814   & 365.7   & 3.047[4]   \\
Sr   & 55.97  & 58.23    &  100.8  & 1354    & 181.8   & 2433   & 442.0   & 4.629[4]  \\
\end{tabular}
\end{ruledtabular}
\end{table*}

The three-body dispersion coefficients for the X-Ne-Ne and
X-Ar-Ar systems are given in Table \ref{XNeNe} while those for the
X-Kr-Kr and X-Xe-Xe systems are given in Table \ref{XKrKr}.  The
quadrupole excitation is associated with the impurity X atom for
the $Z^{\rm XRgRg}_{211}$ coefficient.

\begin{table*}[th]
\caption[]{  \label{XNeNe} The three-body dispersion coefficients
(in atomic units), for systems containing two neon atoms and two
argon atoms.  The sources for the $f$-values distributions are the
same as those in Table \ref{homo}.  Results in the Midzuno-Kihara
(MK) approximation \cite{midzuno56a,kihara58a} use the present
oscillator strength distributions to compute the underlying
$\alpha_d$ and $C_6$ needed as input. The numbers in the square
brackets denote powers of ten. } \vspace{0.1cm}
\begin{ruledtabular}
\begin{tabular}{lcccccccc}
Systems &  \multicolumn{2}{c}{$Z_{111}$} &
          \multicolumn{1}{c}{$Z_{112}$}  &  \multicolumn{1}{c}{$Z_{211}$}  &
          \multicolumn{1}{c}{$Z_{122}$}  &  \multicolumn{1}{c}{$Z_{221}$}  &
          \multicolumn{1}{c}{$Z_{113}$}  &  \multicolumn{1}{c}{$Z_{311}$}  \\
\cline{2-3}
    &     Present & MK  &      \\ \hline
  \multicolumn{9}{c}{X-Ne-Ne} \\
H    & 12.64  & 13.01   &  35.61     & 49.82    & 100.3      & 140.4      & 200.5    & 467.3   \\
Li   & 111.9  & 113.6   &  315.2     & 1907     & 887.9      & 5370       & 1768     & 6.502[4]  \\
Na   & 126.2  & 130.2   &  355.6     & 2567     & 1002       & 6638       & 1994     & 8.629[4]       \\
K    & 191.6  & 200.9   &  539.9     & 4803     & 1521       & 1.353[4]   & 3028     & 2.268[5]   \\
Rb   & 215.1  & 228.0   &  606.0     & 5785     & 1707       & 1.630[4]   & 3399     & 2.893[5]   \\
Be   & 62.79  & 64.37   &  176.9     & 674.8    & 498.3      & 1901       & 994.2    & 1.060[4]     \\
Mg   & 102.5  & 105.6   &  288.8     & 1491     & 813.5      & 4199       & 1622     & 3.156[4]     \\
Ca   & 175.6  & 183.1   &  494.6     & 175.6    & 1393       & 1.073[4]   & 2776     & 1.157[5]   \\
Sr   & 212.2  & 223.0   &  598.0     & 5111     & 1685       & 1.440[4]   & 3356     & 1.754[5]  \\
  \multicolumn{9}{c}{X-Ar-Ar} \\
H    & 174.0  & 177.9   &  846.6     & 658.1    & 4127       & 3217       & 8470     & 6056   \\
Li   & 1818   & 1840    &  8704      & 2.955[4] & 4.170[4]   & 1.420[5]   & 8.804[4] & 9.836[5]  \\
Na   & 2022   & 2068    &  9690      & 3.675[4] & 4.647[4]   & 1.766[5]   & 9.793[4] & 1.314[6]  \\
K    & 3041   & 3169    &  1.459[4]  & 7.636[4] & 7.006[4]   & 3.662[5]   & 1.474[5] & 3.532[6]  \\
Rb   & 3378   & 3557    &  1.622[4]  & 9.220[4] & 7.798[4]   & 4.208[5]   & 1.638[5] & 4.525[6]  \\
Be   & 945.6  & 962.3   &  4557      & 9714     & 2.198[4]   & 4.704[4]   & 4.585[4] & 1.476[5]    \\
Mg   & 1568   & 1600    &  7544      & 2.220[4] & 3.633[4]   & 1.071[5]   & 7.599[4] & 4.548[5]    \\
Ca   & 2719   & 2812    &  1.307[4]  & 5.890[4] & 6.290[4]   & 2.832[5]   & 1.318[5] & 1.732[6]    \\
Sr   & 3279   & 3415    &  1.577[4]  & 7.970[4] & 7.590[4]   & 3.829[5]   & 1.590[5] & 2.655[6]    \\
\end{tabular}
\end{ruledtabular}
\end{table*}

\subsection{Other systems}

The sheer number of possible heteronuclear three-body dispersion
coefficients precludes their listing and discussion in the present
manuscript. The three-body dispersion coefficients for every
possible three-body trimer formed from the atoms studied in this
paper (H, He, Ne, Ar, Kr, Xe, Li, Na, K, Rb, Be, Mg, Ca, Sr) are
given in the supplementary data published with this paper.

Previous compilations of the $Z^{abc}_{111}$ three-body coefficients
have been given for all possible alkali-atom trimers from Li to Cs
\cite{marinescu97a,mitroy03f}.  In addition, $Z^{abc}_{111}$
coefficients for all possible alkaline-earth atoms from Be to Sr
have also been published \cite{mitroy03f}.  The supplementary data
go well beyond these existing tabulations in that the higher-order
$Z^{\rm abc}_{112}$, $Z^{\rm abc}_{113}$, and $Z^{\rm abc}_{122}$
are also given.

\begin{table*}[th]
\caption[]{  \label{XKrKr} The three-body dispersion coefficients
(in atomic units), for systems containing two krypton or two xenon
atoms.  The sources for the $f$-values distributions are the same as
those in Table \ref{homo}.  Coefficients in the Midzuno-Kihara (MK)
approximation \cite{midzuno56a,kihara58a} use the present oscillator
strength distributions to compute the underlying $\alpha_d$ and
$C_6$ needed as input.  All values are in atomic units. The numbers
in the square brackets denote powers of ten. } \vspace{0.1cm}
\begin{ruledtabular}
\begin{tabular}{lcccccccc}
Systems &  \multicolumn{2}{c}{$Z_{111}$} &
          \multicolumn{1}{c}{$Z_{112}$}  &  \multicolumn{1}{c}{$Z_{211}$}  &
          \multicolumn{1}{c}{$Z_{122}$}  &  \multicolumn{1}{c}{$Z_{221}$}  &
          \multicolumn{1}{c}{$Z_{113}$}  &  \multicolumn{1}{c}{$Z_{311}$}  \\
\cline{2-3}
    &     Present & MK  &      \\ \hline
  \multicolumn{9}{c}{X-Kr-Kr} \\
H    & 370.1     & 380      & 2485     & 1385     & 1.352[4]   & 8409     & 2.760[4] & 1.268[4] \\
Li   & 4081      & 4136     & 2.407[4] & 6.612[4] & 1.420[5]   & 3.870[5] & 3.044[5] & 2.151[6]  \\
Na   & 4521      & 4628     & 2.671[4] & 8.216[4] & 1.579[5]   & 4.819[5] & 3.372[5] & 2.881[6]  \\
K    & 6795      & 7.086    & 4.017[4] & 1.700[5] & 2.377[5]   & 1.006[6] & 5.070[5] & 7.800[6]   \\
Rb   & 7526      & 7934     & 4.454[4] & 2.055[5] & 2.639[5]   & 1.215[6] & 5.616[5] & 1.000[7]  \\
Be   & 2066      & 2109     & 1.232[4] & 2.092[4] & 7.354[4]   & 1.256[5] & 1.539[5] & 3.147[5]    \\
Mg   & 3445      & 3523     & 2.049[4] & 4.831[4] & 1.220[5]   & 2.885[5] & 2.567[5] & 9.796[5]    \\
Ca   & 6015      & 6229     & 3.569[4] & 1.299[5] & 2.120[5]   & 7.714[5] & 4.484[5] & 3.774[6]    \\
Sr   & 7253      & 7563     & 4.304[4] & 1.763[5] & 2.557[5]   & 1.046[6] & 5.407[5] & 5.807[6]    \\
   \multicolumn{9}{c}{X-Xe-Xe} \\
H    & 871.4     & 899.3    & 7139     & 3217     & 5.862[4]   & 2.650[4] & 1.139[5] & 2.930[4]  \\
Li   & 1.034[4]  & 1.049[4] & 8.232[4] & 1.610[5] & 6.558[5]   & 1.294[6] & 1.326[6] & 5.255[6]  \\
Na   & 1.140[4]  & 1.168[4] & 9.092[4] & 2.013[5] & 7.257[5]   & 1.616[6] & 1.464[6] & 7.059[6]  \\
K    & 1.713[4]  & 1.787[4] & 1.366[5] & 4.528[5] & 1.091[6]   & 3.405[6] & 2.200[6] & 4.258[7]  \\
Rb   & 1.891[4]  & 1.994[4] & 1.510[5] & 5.157[5] & 1.208[6]   & 4.120[6] & 2.430[6] & 2.483[7]  \\
Be   & 5042      & 5162     & 4.072[4] & 5.009[4] & 3.293[5]   & 4.076[5] & 6.531[5] & 7.439[5]    \\
Mg   & 8476      & 8690     & 6.824[4] & 1.172[5] & 5.500[5]   & 9.489[5] & 1.095[6] & 2.345[6]    \\
Ca   & 1.494[4]  & 1.549[4] & 1.199[4] & 3.210[5] & 9.625[5]   & 2.580[6] & 1.962[6] & 9.177[6]  \\
Sr   & 1.802[4]  & 1.881[4] & 1.445[5] & 4.377[5] & 1.161[6]   & 3.512[6] & 2.323[6] & 1.419[7]  \\
\end{tabular}
\end{ruledtabular}
\end{table*}

\subsection{The Midzuno-Kihara approximation}

The Midzuno-Kihara (MK) approximation \cite{midzuno56a,kihara58a}
for $Z_{111}$ has been given for all combinations in Tables
\ref{homo}, \ref{XHeHe} \ref{XNeNe} and \ref{XKrKr}. The
coefficients in the MK approximation use the present oscillator
strength distributions to compute the underlying $\alpha_d$ and
$C_6$ needed as input.   In every instance the MK results are larger
by an amount between 1-4$\%$ than those of the explicit calculation.

\section{Conclusions}

A large scale investigation of the three-body dispersion
coefficients for a set of 14 atoms has been completed.  The results
presented for the set of small systems, namely H, He and Li should
be regarded as the existing calculational standard.

The current set of values should also be adopted as the reference
standard to be used for heavier systems.  Even though more
approximate methods of solving the Schr\"odinger equation are
adopted, the semi-empirical Hamiltonian adopted for the alkali atoms
and the alkaline earth atoms reduces the computational complexity to
such an extent that uncertainties in the solution of the resulting
Schr\"odinger equation are largely eliminated.  When comparisons of
polarizabilities and two-body dispersion coefficients have been made
the CICP approach has typically been within 1$\%$ of the most
sophisticated approaches to atomic structure available
\cite{mitroy03f,porsev03a}.

Besides an improvement in accuracy, the scope of the present work
exceeds that of any previous work by at least an order of magnitude.
For example, Huang and Sun \cite{huang11a}, and Patil and Tang
\cite{patil97a}. presented extensive results for the alkali atoms.
However, the present calculations also encompass the rare gases and
the alkaline-earth atoms.

\begin{acknowledgments}
This work was supported by NNSF of China under Grant Nos. 11104323
and 11034009, and by the National Basic Research Program of China
under Grant No. 2012CB821305. Z.-C.Y. was supported by NSERC of
Canada and by the computing facilities of ACEnet, SHARCnet,
WestGrid, and in part by the CAS/SAFEA International Partnership
Program for Creative Research Teams. J. M. would like to thank the
Wuhan Institute of Physics and Mathematics for its hospitality
during his visits. The work of J.M was supported in part by the
Australian Research Council Discovery Project DP-1092620. ITAMP is
partially supported by a grant from the US NSF to Harvard University
and the Smithsonian Astrophysical Observatory.
\end{acknowledgments}


\end{document}